\documentclass[pre,reprint,superscriptaddress]{revtex4-1}
\usepackage{subfigure}
\usepackage{graphicx}
\usepackage{epsfig}
\usepackage{amsmath}
\usepackage{epstopdf} 
\usepackage{color}
\usepackage{mathtools}
\usepackage{float}

\usepackage{hyphenat}
\usepackage[colorlinks=true,linkcolor=blue,urlcolor=blue,citecolor=blue]{hyperref}

\newcommand{\bracketed}[1]{\left[\vphantom{\let\\=\relax #1}\right. #1 \left.\vphantom{\let\\=\relax #1}\right]}

\begin{document}
\title{Adsorption of melting deoxyribonucleic acid}

\author{Debjyoti Majumdar}
\email{debjyoti@post.bgu.ac.il}
\affiliation{Alexandre Yersin Department of Solar Energy and Environmental Physics, Jacob Blaustein Institutes for Desert Research,\\
Ben-Gurion University of the Negev, Sede Boqer Campus 84990, Israel}
\date{\today}

\begin{abstract}
The melting of a homopolymer double-stranded (ds) deoxyribonucleic acid (DNA) in the dilute limit is studied numerically in the presence of an attractive and impenetrable surface on a simple cubic lattice. The two strands of the DNA are modeled using two self-avoiding walks, capable of interacting at complementary sites, thereby mimicking the base pairing. The impenetrable surface is modeled by restricting the DNA configurations at the $z\geq 0$ plane, with attractive interactions for monomers at $z=0$. Further, we consider two variants for $z=0$ occupations by ds segments, where one or two surface interactions are counted. This consideration has significant consequences, to the extent of changing the stability of the bound phase in the adsorbed state. Interestingly, adsorption changes from critical to first-order with a modified exponent on coinciding with the melting transition. For simulations, we use the pruned and enriched Rosenbluth algorithm.	  
\end{abstract}

\maketitle

\section{Introduction} 

The denaturation of the double-stranded (ds) deoxyribonucleic acid (DNA) from a bound (ds) to an unbound single-stranded (ss) phase is an important step towards fundamental biological processes such as DNA replication, ribonucleic acid (RNA) transcription, packaging of DNA and repairing \cite{watson2003}. {\it In vitro}, the melting transition is induced by changing the temperature or {\it p}H of the DNA solution. However, the physiological condition would allow neither extremes of temperature nor {\it p}H level inside the cell.  Therefore, the cell has to rely on other ambient factors to locally modify the stability of the ds structure of the DNA. Among others, one of the crucial factors and a potential candidate that can alter the stability of the native DNA form is an interaction of the DNA with a surface, e.g., in the form of proteins or cell membranes. The DNA strands being polymers, can undergo an adsorption transition, where the two strands, either in the ds or ss phase, get adsorbed on a surface \cite{eisen1982}. {\it In vivo,} the protein-induced DNA-membrane complex is used during the replication process, cell division, and for inducing local bends in the rigid duplex DNA \cite{firshein1989,kapri2008}. Again, adsorption is instrumental in packaging DNA inside the virus heads \cite{carri1999, purohit2005}. On the technological front, the adsorbing property of the DNA is often used to target drug delivery in gene therapy \cite{bathaie1999,radler1997}, and for manufacturing biosensors with quick and accurate detection of DNA in bodily samples. In all these instances, the surface-DNA interaction can be tuned by changing the nature of the surface. This tunability calls for a detailed phase mapping arising from the interaction of the DNA with the adsorbing surface.

The melting and the adsorption transition individually form the subject of many theoretical and experimental studies in the past. However, studies investigating the melting-adsorption interplay remains relatively less {explored}. Naively, one would expect four distinct phases when melting and adsorption are considered together \cite{kapri2008}. However, the unbound-adsorbed phase was found missing in a theoretical study \cite{verdyan2006}, which employs an exactly solvable model of flexible, ideal chains. Overall, in Ref.~\cite{verdyan2006}, it was found that the bound state is stabilized in the presence of an adsorbing surface. By contrast, on the experimental side, Ref.~\cite{schreiner2011} had demonstrated that directly adsorbed DNA hybrids are significantly less stable than if free. Therefore, further study of the melting-adsorption interplay employing more versatile models is essential for a complete understanding.

Numerically,  lattice models have been helpful in extracting sensible results on par with the experiments, e.g.,  the melting transition  was shown to be first-order when excluded volume interactions are fully included \cite{causo2000}. In contrast, the polymer adsorption transition was shown to be continuous \cite{eisen1982,grassberger2005}. It is, therefore, instructive to include excluded volume interaction while constructing a model for DNA melting. With this in mind, in this paper, we explore the interplay between the melting and the adsorption transitions of a model homopolymer DNA, using a lattice adaptation of the Poland-Scheraga model on a simple cubic lattice where self-avoidance can be duly implemented among the intra- and inter-strand segments \cite{causo2000}. 

Further, we consider two model variants depending on how the ds segments interact with the surface. We found that the melting vs. adsorption phase diagram is drastically different for the two different {interaction schemes} between the ds and the adsorbing surface. {In one of the models,} the two transitions  coalesce into a single transition {for specific values of the coupling potentials,} thereby promoting the continuous adsorption transition to first-order. However, the first-order nature of the melting transition remains unaffected in both cases even when there is a change in the dimensionality.

The remaining paper is organized in the following manner. In Sec.~\ref{sc2} we introduce the DNA-surface interaction model and its two variants. In Sec.~\ref{sc3} we describe the simulation algorithm required to generate the equilibrium configurations over an adsorbing surface. In Sec.~\ref{sc4} we qualitatively describe the problem along with the thermodynamic observables required to study the problem. Next, we discuss the findings of model I in Sec.~\ref{sc5}(A) and model II in Sec.~\ref{sc5}(B). Finally, we conclude the paper in Sec.~\ref{sc6}.

\begin{figure}
\centering
\includegraphics[width=\linewidth, angle=0]{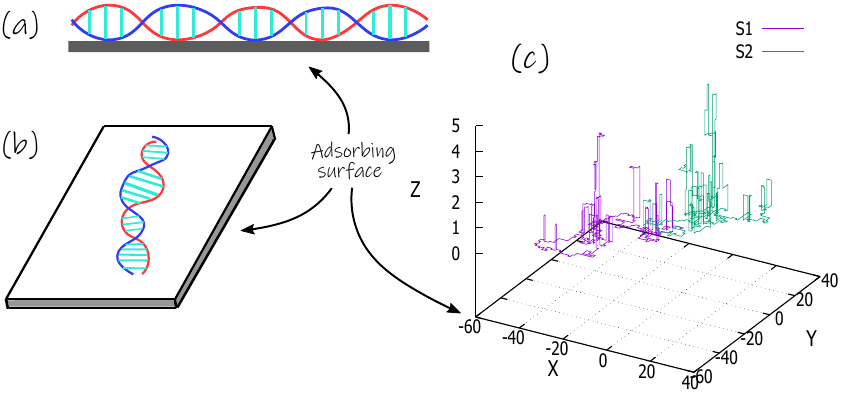}
\caption{(Color online) Schematic diagram for (a) lateral view of model I, and (b) planar view of model II. In (a) representing model I, only one strand is effectively interacting with the surface in the bound state. While both the strands are simultaneously in contact with the surface in model II as in (b). (c) {A typical configuration showing strand A (S1) and strand B (S2) with the adsorbing plane at z = 0}.}
\label{fig1}
\end{figure}

\section{The model} \label{sc2}
We model the DNA strands (say A and B) as two self-avoiding walks (SAWs), represented by the vectors ${\bf r}^A_i$ and ${\bf r}^B_j$ ($1\leq i,j \leq N$). {The strands are also mutually avoiding, with the exception that they are} capable of forming a base pair (bp) among the complementary monomers ($i=j$) from the two strands while occupying the same lattice site (${\bf r}^A_i={\bf r}^B_i$). One end of the DNA is grafted in the $z=0$ plane. The other end is free to wander in the $z\geq 0$ direction, with the $z=0$ plane impenetrable and attractive. An energy $-\epsilon_{bp}$ is associated with each bound bp independent of the bp index (homopolymer) and is represented by the reduced variable $g=\epsilon_{bp}/k_BT$, where $T$ is the temperature, and $k_B$ is the Boltzmann constant. For each interaction with the $z=0$ surface, there is an energetic gain of $-\epsilon_s$, represented by the reduced variable $q=\epsilon_s/k_BT$. Further, we consider two variants: model I and model II. The difference in the two variants is in the strength of the ds interaction with the surface {($\epsilon_{s\text{-}ds}$)}; in model I, we consider only one unit of interaction (${\epsilon_{s\text{-}ds}}=\epsilon_s$), while in model II, we consider two units of interaction (${\epsilon_{s\text{-}ds}}=2\epsilon_s$), each for one of the strands. The rationale behind such speculation is that when interacting sidewise, like in Fig.~\ref{fig1}(a), there would be an effective interaction of one strand. By contrast, when both the strands touch the plane simultaneously, each strand would contribute [Fig.~\ref{fig1}(b)]. These two scenarios may arise depending on the hardness of the surface. While metallic surfaces (such as Gold) used during experiments are hard, biological surfaces tend to be much softer. {For ss segments, however, we always consider only one unit of surface interaction ($\epsilon_{s\text{-}ss}=\epsilon_s$)}. A {typical configuration from our simulation} is shown in Fig.~\ref{fig1}(c). The Hamiltonian for a typical configuration according to model II can be written as,
\begin{equation}
\beta\mathcal{H} = - g\sum_{i=1}^N \delta_{{\bf r}^A_i, {\bf r}^B_i} -q\sum_{i=1}^N\sum_{\alpha=A,B} \delta_{0,z^{\alpha}_i}
\end{equation}
where $\beta=1/(k_BT)$ and $\delta_{i,j}$ is the Kronecker delta. We set the Boltzmann constant $k_B=1$ throughout our study. 

The two variants discussed above only represent the two exclusive scenarios. However, in reality we might have something in-between where both of the above pictures hold simultaneously at different places along the DNA. Again, the adsorbing surface can generally be of complex geometry with varying degree of roughness and curvature. However, we choose a smooth and impenetrable flat surface for simplicity. Other than that, for simplicity, and due to limitations of the considered model, we neglect some aspects of the DNA, such as its heterogeneous sequence, the difference in rigidity of the ss and ds segments, helical geometry, etc. We aim to consider some of these in our future work. The above-described DNA model, despite being minimalistic, has been successfully used to predict the first-order nature of the melting transition \cite{causo2000}, change of DNA rigidity around the melting transition \cite{majumdar2020}, the unzipping and stretching transitions \cite{majumdar2021}, and a non-monotonic change of the melting temperature in a poor solvent background \cite{majumdar2022}.

\section{Simulation Methods} \label{sc3}

We use the pruned and enriched Rosenbluth algorithm (PERM) \cite{grassberger1997} to simulate the equilibrium configurations of the dsDNA over an attractive surface. Two strands are grown at once, adding monomers on top of both strands' last added monomer at once.  At each step, we calculate the joint possibilities of stepping into free sites obtained by a Cartesian product of the individual sets of possibilities, i.e., $\mathcal{S}_n=\mathcal{S}_n(A)\times \mathcal{S}_n(B)$. Each element in {$\mathcal{S}_n$} corresponds to an ordered pair of new steps for both the strands and carries a Boltzmann weight of $\exp(g\times l+q\times m)$, where $l=1$ for a bp, and $0$ otherwise, while {$m=0,1$ or $2$} depending upon the number of surface contacts and model. Then, a choice is made from the {$\mathcal{S}_n$} set of possibilities according to the {\it importance sampling}. At each step, the local partition function is calculated as $w_n=\sum_{\mathcal{S}_n}\exp(g\times l+q\times m)$. The partition sum {at} length $n$ \cite{comm3} is then estimated by product over the local partition sums at each step, $W_n=\prod_{i=1}^n w_i$, and averaging over the number of started tours {we obtain} $Z_n = \langle W_n \rangle $. Enrichment and pruning at $n$th step is performed depending on the ratio, {$r=W_n/Z_n$} and using the scheme:

\begin{equation}
r=\begin{cases}
1, & \text{continue to grow} \\
<1, & \text{prune with probability $(1-r)$} \\
>1, & \text{make $k$-copies.}
\end{cases}
\end{equation}

If $r<1$ and pruning fails, the configuration is continued to grow but with $W_n=Z_n$. For enrichment ($r>1$) $k$ is chosen as, $k=min(\lfloor r \rfloor,\mathcal{N}(\mathcal{S}_n))$, where $\mathcal{N}(\mathcal{S}_n)$ is the cardinality of the set $\mathcal{S}_n$, and each copy carries a weight $\frac{W_n}{k}$ . {A depth-first approach is employed for creating the copies, where the configuration of a single copy is dealt with at a time, and recursion is used to start the different copies from the same enrichment point.} Averages are taken over $10^8$ tours. 

To estimate {averages of} thermodynamic observables (say $Q_n$) at length $n$,  the averaging is performed on the fly using the expression:
\begin{equation}
\label{SMEQ1}
\langle Q_n \rangle (g,q) = \frac{\langle Q_n W_n(g,q) \rangle}{Z_n(g,q)}
\end{equation}

where the $\langle \cdots \rangle$ in the numerator represents the {ensemble} average of the quantity over the number of started tours, using the local estimate of the configuration weight $W_n$.

One of the important aspects in simulating lattice self-avoiding walks is in checking if the immediate next sites are empty. The straightforward way is to check if any of the last $N-1$ steps occupy the site. However, for walks of length $N$ the time required in this operation grows as $\mathcal O (N)$, and $\mathcal O (N^2)$ for the total chain. This can be avoided using the {\it bit map} method in which the whole lattice is stored in an array using a hashing scheme where each site is given an array address like: $f(x,y,z)=x+yL+zL^2+offset$, where $L$ is an odd number, representing the dimension of the virtual lattice box and $offset=\lfloor L^d/2 \rfloor$ is a constant number which depends upon $L$ to make the address start from zero. Here, the checking of self-avoidance is  $\approx \mathcal O (1)$, with no possibility of {\it hashing collision}. However, since our problem requires constraining the polymer above the plane on which it is grafted, there is a significant chance that the polymer will move out of the simulation box. A lower bound on the linear box dimension {demands} that $L\gg 2N^{\nu_{2d}}$, where the factor $2$ considers the chain starts growing from the center of the plane and $\nu_{2d}$ is the SAW size exponent in two-dimensions.  The amount of required memory increases rapidly with $L$.

A possible way out is to use a {\it linked list} method, e.g., the {Adelson-Velsky-Landis} (\textsc{AVL})  {binary search tree} \cite{avl_tree}. The \textsc{AVL} algorithm creates a tree-like structure where each node represents an occupied lattice site. Each entry for a new step is associated with \emph{search}, \emph{insertion}, and \emph{rebalancing} the tree branches. Each \emph{insertion} or \emph{deletion} operation requires $\mathcal O (\log(n))$ time, where $n$ is the total number of nodes which translates to the number of monomers or occupied sites or the polymer length. For a chain of length $N+1$, the total growth time (assuming only \emph{insertion} is performed) is: $\ln(1)+\ln(2)\cdots \ln(N)=\ln(N!)$. Using Sterling approximation, and for large $N$, this is approximately $\mathcal{O}(N\ln N)$. Moreover, the \textsc{AVL} algorithm can be easily incorporated into the recursive structure of the \textsc{PERM} algorithm.  

\section{Qualitative description and quantities of interest}  \label{sc4}
Before describing the findings of our study, let us briefly discuss a few of the results known so far, along with the thermodynamic quantities we would be interested in. The melting of the dsDNA with excluded volume interaction is a first-order transition \cite{causo2000}. The bound and unbound phases are dominated by energy and entropy, respectively, depending upon whichever minimizes the free energy. The average number of bound bps per unit length ($n_c/N$) serves as the order parameter with $n_c/N=1$ and $0$ in the bound and unbound phase, respectively. The fluctuation in $n_c$ is denoted by $C_c=\langle n_c^2\rangle -\langle n_c\rangle^2$, and the associated crossover exponent as $\phi_m$. For first-order melting transition $\phi_m=1$, and $\phi_m<1$ for continuous melting transition. For this specific model, in the absence of any adsorbing surface (i.e., $q=0$), the melting takes place at $g^*=1.3413$ with the crossover exponent $\phi_m=0.94$, which is close to one, as expected for a first-order transition  \cite{causo2000}. On the other hand, the 3d to 2d adsorption of a lattice polymer on a two-dimensional surface is a continuous transition with the critical point at $q_c=0.2856$ \cite{grassberger2005}. Here, the average number of surface contacts per unit length $n_s/N$ is the order parameter, and its fluctuation is denoted by $C_s$. The corresponding critical exponent controlling the growth of surface contacts at the critical point is $\phi_a$. The exponent $\phi_a$ is expected to be universal, with a value of $\phi_a=1/2$ from mean-field calculations at the critical point.  However, from computer simulations, the most recent improved estimate of the critical exponent suggest $\phi_{a}=0.48(4)$ \cite{grassberger2005,bradly2018}. Interestingly, one can visualize the naturation of a ds DNA  as selective adsorption of one of the strands on the other, with the surface being one-dimensional and fluctuating. Then, if we reduce the fluctuation by pulling the strands, preferably in the same direction, the melting transition indeed becomes continuous \cite{majumdar2021}.

Often, systems undergoing multiple transitions (i.e., described by multiple order parameters) may result in a m\'elange of critical exponents obtained from different methods such as the finite-size-scaling analysis, scaling of the specific heat peaks with the system size, the reunion exponent also known as the bubble-size-exponent for DNA, among others. Therefore, deciding the behavior of the transition becomes difficult. In these situations, to corroborate any change in the nature of the transition, the general prescription is to look at the probability distribution of the associated order parameter close to the transition point. For adsorption transition, we look at the probability distribution $(P_{n_s})$ of the surface contacts ($n_s$) at different lengths, {close to the} transition point ($q_c$). To calculate $P_{n_s}(q,g)$, we find the conditional partition sum $Z_{n,n_s}(q,g)$, where $n$ is the {length of the DNA} having $n_s$ number of surface contacts. Finally, $P_{n_s}$ is found using the formula,
\begin{equation}
\label{SMEQ2}
P_{n_s}(q,g) = \frac{Z_{n,n_s}(q,g)}{\sum_{n_s=0}^{2n} Z_{n,n_s}(q,g)}
\end{equation}
where the maximum number of surface contacts {to sum over} is always double the number of possible bps {($n$), which happens} for the unbound configuration. For a continuous transition, the order parameter distribution is expected to hold a scaling relation of the form 
\begin{equation}
P_{n_s}\sim N^{-\phi} p(n_s/N^{\phi})
\label{SMEQ9}
\end{equation}
where $\phi$ is the associated scaling exponent. It is worthwhile to note that  Eq. \ref{SMEQ9} is also true for the first-order melting transition in our DNA model \cite{causo2000}.

For $q<q_c$, the partition sum of a SAW scales as
\begin{equation}
\label{SMEQ3}
Z(q,N) \sim \mu^N N^{\gamma_1 - 1}
\end{equation}
where the subscript $1$ in the entropic exponent $\gamma_1$ denotes the fact that one end is grafted on an impenetrable surface, while the exponential growth through $\mu$ (the {\it effective coordination} number) is invariant. Near the adsorption transition ($q\sim q_c$), $Z(q,N)$ should scale as
\begin{equation}
\label{SMEQ4}
Z(q,N)\sim \mu^N N^{\gamma_1'-1} \psi[(q-q_c)N^{\phi_a}]
\end{equation}
where $\psi(x)$ is a scaling function. Taking derivative of $\ln Z(q,N)$ in Eq.~(\ref{SMEQ4}) with respect to $q$, and setting $q=q_c$, one obtains the scaling form of the mean adsorbed energy at the critical point as

\begin{equation}
\label{SMEQ5}
n_s \sim N^{\phi_a}.
\end{equation}

Therefore, at the critical adsorption point, the quantity {$n_s/N^{\phi_a}$} should be $N$ independent for appropriate $\phi_a$ in the thermodynamic limit $N\rightarrow \infty$. We will use this quantity to estimate the critical point for adsorption, where for continuous adsorption transitions, we use $\phi_a=1/2$.  

Further, following Ref.~\cite{grassberger2005}, we also looked at the quantity,

\begin{equation}
\label{SMEQ6}
\gamma_{1,eff}' = 1 + \frac{\ln \left[Z(q,2N)/Z(q,N/2)/\mu^{3N/2} \right]}{\ln 4}
\end{equation}

using $\mu=4.6840386$. We simulate chains of length up to $N=10,000$, to extract data  up to $N=5000$ using Eqs. \ref{SMEQ5} and \ref{SMEQ6}. 

For melting, we estimated the transition points from the average number of bound bps ($n_c$) and its fluctuation $(C_c)$. The melting points are obtained from the scaling of $n_c$ and $C_c$, following the equations,
\begin{equation}
\label{SMEQ7}
n_c\sim N^{\phi_m} f[(g-g^*)N^{\phi_m}]
\end{equation}
and,
\begin{equation}
\label{SMEQ8}
C_c\sim N^{2\phi_m} h[(g-g^*)N^{\phi_m}].
\end{equation}

Tuning $g^*$ and $\phi_m$ to the appropriate values in Eq.~(\ref{SMEQ7}) and (\ref{SMEQ8}) would make the data for different lengths fall upon each other resulting in {\it data collapse}.

Finally, apart from Eq.~\ref{SMEQ5}, we also use the crossing point of the $C_s$ curves of the two longest lengths to {locate} the critical point for the continuous adsorption transitions. However, for first-order adsorption, the method of data collapse is used using Eq.~(\ref{SMEQ7}) and (\ref{SMEQ8}) but with $q$ in place of $g$, and, $n_c$ and $C_c$ replaced with $n_s$ and $C_s$, respectively. Moreover, we can have an idea about the nature of the transition and the transition point beforehand from the shape of the $C_s$ curves.

\section{Results and discussions} \label{sc5}

\subsection{Model I} \label{sc5A}
\begin{figure}[t]
\centering
\includegraphics[width=\linewidth]{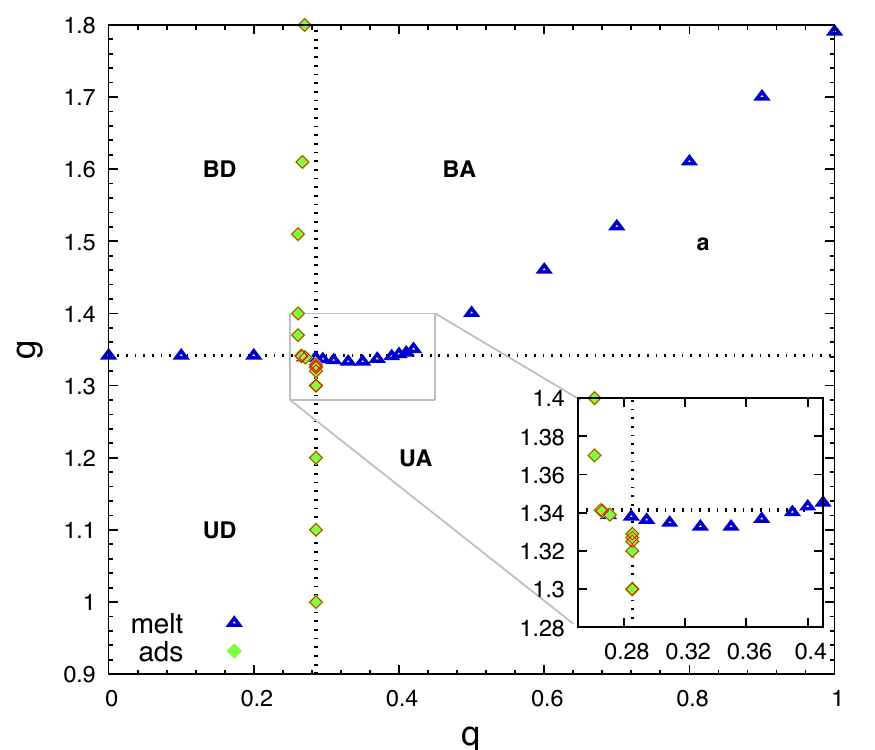}
\caption{(Color online) Model I phase diagram for melting `melt' and adsorption `ads' . The dotted lines represent the transition points for the individual cases; for melting $g^*=1.3413$, and  $q_c=0.2856$ for adsorption. (Inset) A scale up of the part of the phase diagram showing a decrease in the threshold $g^*$ for bound state.  The error bars in $q_c$ and $g^*$ are of the size of the plotting points.}
\label{fig2}
\end{figure}

We plot the melting vs. adsorption phase diagram for model I in Fig.~\ref{fig2}. For reference to the pure cases, the two individual transitions, $g^*=1.3413$ for the first-order melting and $q_c=0.2856$ for the continuous adsorption transition, are plotted using the dotted lines. {However, when both are present,} as we change the parameters, these two lines cross at a multicritical point somewhere around $g\approx 1.34$ and $q\approx 0.265$, {thereby}, dividing the phase plane into four equilibrium phases,  viz., bound-desorbed ({\bf BD}), unbound-desorbed ({\bf UD}), unbound-adsorbed ({\bf UA}), and the bound-adsorbed ({\bf BA}) phase. These new mixed phases emerge as a result of coupling between the melting and adsorption transitions, e.g., region {\bf a} {in Fig.~\ref{fig2}} corresponds to an unbound phase which otherwise should have been bound. 
  
 As the two lines ($g^*=1.3413$ and $q_c=0.2856$) {intersect} each other, the bound state is primarily stabilized for increasing $q$, which is somewhat surprising [see Fig. \ref{fig2} inset]. {Also, the critical adsorption line slightly deviates from the pure value around the junction}. The increased stability of the bound state persists for a small range of $q$ values $(0.26(6)\lesssim q\lesssim 0.4)$. It is because, in this region, the bound and unbound phases in the vicinity of the melting line are unequally placed in their corresponding adsorbed phases. With the ds phase placed relatively deeper into the adsorbed phase is entropically stabilized, {owing to the loss in entropy of the adsorbed ds segments}.  This short period of stability is followed by a monotonic increase in the threshold $g^*$ required to achieve a bound state for $q>0.4$, separating the destabilized bound and unbound state in the adsorbed phase. {We found} a linear dependence of $g^*$ on $q$, and fitting the relation $g^*(q)=aq+b$ gives a slope of $a=0.731$.  
\begin{figure}
\centering
\hspace{-7cm}(a)\\
\includegraphics[width=\linewidth]{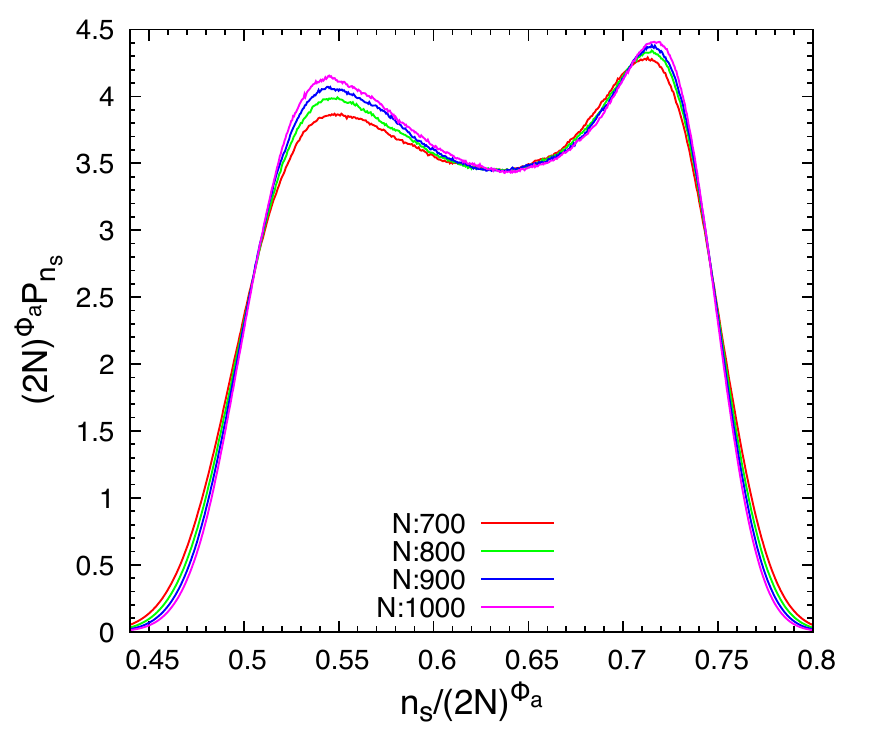}\\
\hspace{-7cm}(b)\\
 \includegraphics[width=\linewidth]{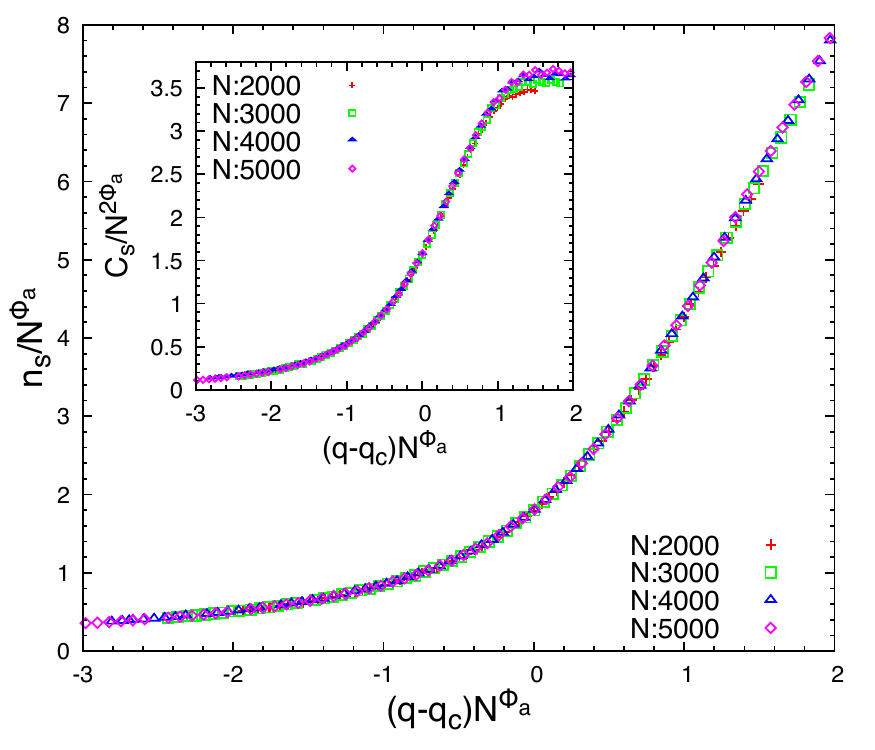}
\caption{(Color online) (a) Model I scaling plot of the the probability distribution $(P_{n_s})$ of surface contacts ($n_s$) on the {\bf BA} $\rightarrow$ {\bf UA} transition line corresponding to the points $g^*=1.5$ and $q=0.659$, {using $\phi_a=1$}. Data shown for chain lengths $N=700$ to $1500$. (b) Scaling plot of the average number of surface contacts $(n_s)$, and (inset) its  fluctuation $(C_s)$ for $g=5$, using $q_c=0.2855$ and $\phi_a=1/2$ for data collapse of chain lengths $N=2000$ to $5000$. }
\label{fig3}
\end{figure}
We can understand this {monotonic increase} using the energy-entropy argument; since the number of independent surface contacts increases upon unbinding, with each ds bp resulting in two new possible ss surface contacts, along with an increase in the entropy, the {\bf UA} phase is strongly favored over the {\bf BA} phase.  A significant consequence is that, the melting in the adsorbed phase ({\bf BA}$\rightarrow${\bf UA}) is different from the pure melting in two-dimensions (2d) where the melting point is at  $g^*_{2d}=0.752(3)$. The disparity {is} a result of the energetic advantage of the {\bf UA} phase over the {\bf BA} phase.
  
   Noticeably, while undergoing {\bf UA} to {\bf BA} transition by varying $q$, the system shows first-order like fluctuation of surface contacts ($C_s$), with the average number of surface contacts $(n_s)$ reducing to half its value than that in the {\bf UA} phase. This observation is supported by the scaling plot of the surface contact probability distribution ($P_{n_s}$) at a point $g^*=1.5$ and $q=0.659$ above the melting phase boundary, using the scaling exponent $\phi_a=0.99$ and Eq. \ref{SMEQ9} [Fig.~\ref{fig3}(a)]. Notice the lower peak {corresponding to {\bf BA} phase} at $n_s/2N\sim 0.5$. However, it is not a genuine desorption transition. It is because for model I the ds and ss surface contacts are treated on equal footing {with equal energetic contribution}. 

For higher $g$ values, the {\bf BA} phase undergoes continuous desorption around $\lim_{g\rightarrow \infty} q_c=0.2856$. In Fig. \ref{fig3}(b), we plot the scaled form of the average surface contacts $n_s$ and its fluctuation $C_s$ in Fig. \ref{fig3}(b) inset  for $g=5$, using the values $q_c=0.285(5)$ and $\phi_a=1/2$ for data collapse.  In the limit $g\rightarrow \infty$, the critical point for adsorption converges to the pure adsorption value since in the completely bound phase ($n_c/N=1$), the adsorption energy per unit length remains the same as that of the unbound case for this model variant.

Summarizing the results of model I, we see that the bound phase is stabilized only for a small range of $q$ values [Fig.~\ref{fig2} inset]. Otherwise, the bound state remains destabilized. For $q<0.265(5)$, the two transitions remain decoupled without affecting each other, while threshold $g^*$ changes rapidly for $q>0.4$, with a linear dependence on $q$. However, the adsorption line has no substantial change except for a small deviation near the crossing point.  

Results involving model I is in accordance with Ref.~\cite{schreiner2011}, where adsorbed DNA hybrids are found to be less stable than their free counterpart. Importantly, these results suggest that since the destabilization of the dsDNA is essential for the ease of opening up a bound segment, adsorption could play a crucial role in initiating certain biological processes related to the transfer of genetic information.

\subsection{Model II}  \label{sc5B}
\begin{figure}[t]
\centering
\includegraphics[width=\linewidth]{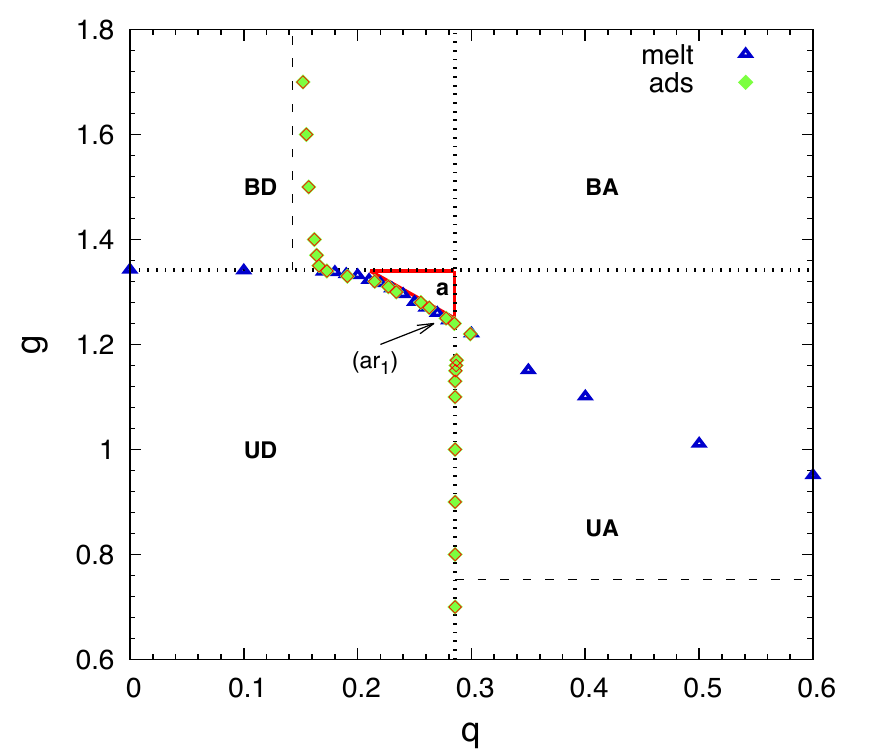}
\caption{(Color online) Model II phase diagram. Dotted lines represent, $g^*=1.3413$ and $q_c=0.2856$. Dashed lines represent, $g^*_{2d}=0.752(3)$ and $q_c=0.1428$. Region {\bf a} corresponds to the Borromean phase \cite{verdyan2006}.}
\label{fig4}
\end{figure}

\begin{figure}
\centering
\hspace{-7cm}(a)\\
\includegraphics[width=\linewidth]{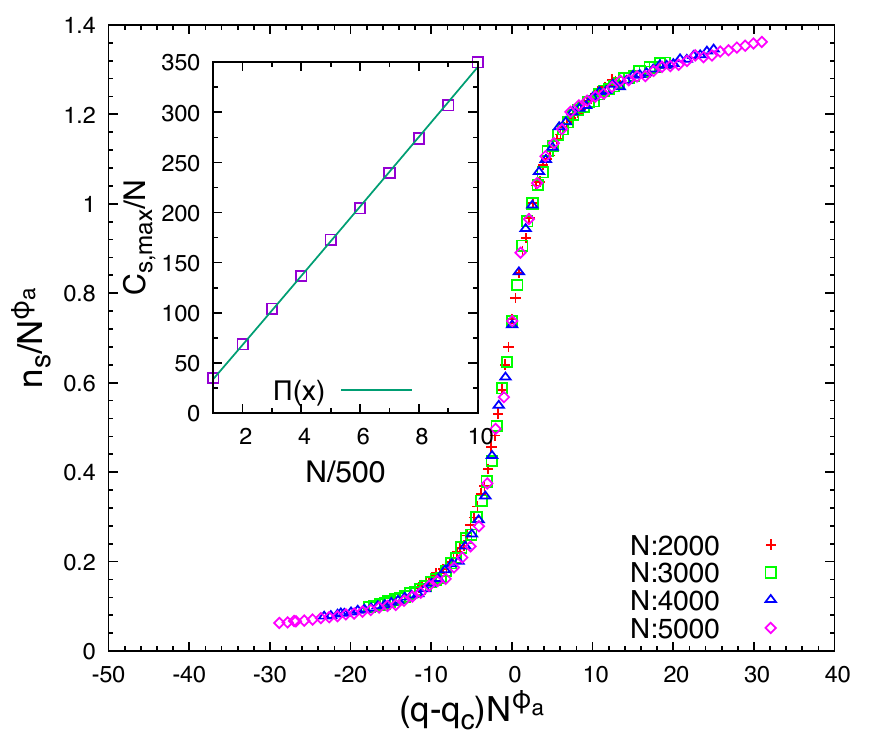}\\
\hspace{-7cm}(b)\\
\includegraphics[width=\linewidth]{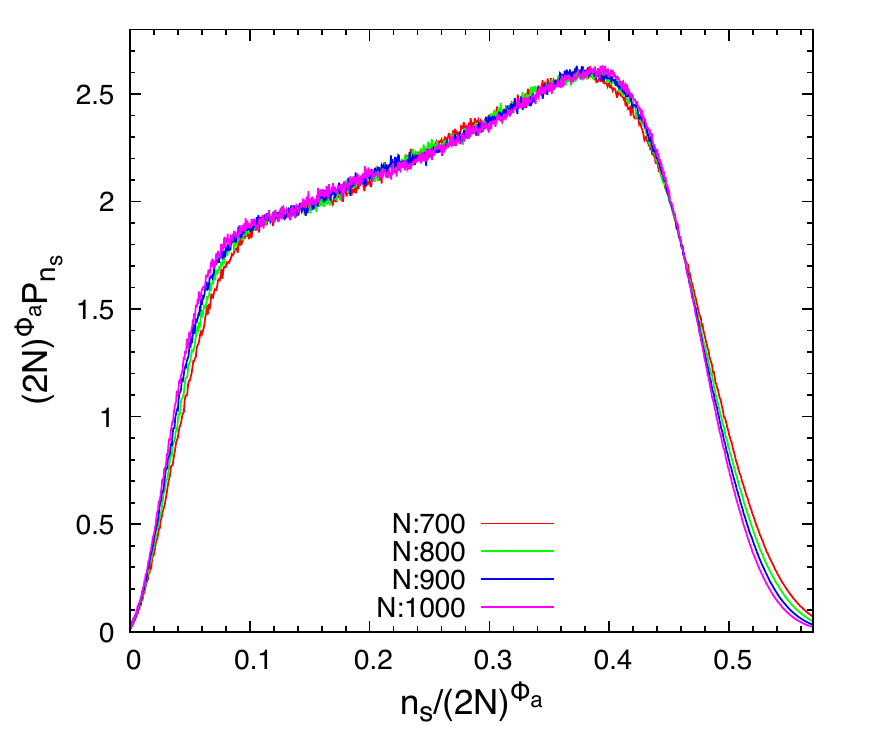}
\caption{(Color online) (a) Model II scaling plot of the average number of surface contacts $n_s$ according to Eq. \ref{SMEQ5}, for $g=1.25$, using $\phi_a=0.9{6\pm 0.02}$ and $q_c=0.2784{\pm 0.0002}$. ((a) inset)  Scaling of the fluctuation peaks $C_{s,max}$ with $N$. $\Pi(x)\sim x^{1.009\pm0.01}$ is a fit to the data points resulting in $\phi_a=1$. (b) Scaled probability distribution of the surface contacts $(P_{n_s})$ at $g=1.25$ and $q=0.2782$,  {using $\phi_a=1$} (arrow $\text{ar}_1$ in {Fig.~\ref{fig4}}). Data shown for chain lengths $N=700$ to $1000$. }
\label{fig5}
\end{figure}

For model II, we consider ds bound segments to have a higher energy gain (precisely, double) than ss segments upon interaction with the surface. Using this scheme of interaction, the phase plane is divided into four distinct phases viz., {\bf BD}, {\bf UD}, {\bf UA} and the {\bf BA} phase [Fig.~\ref{fig4}]. We can further identify three types of melting transition using these four phases: (i) when both the phases are desorbed, (ii) when the bound phase is adsorbed, and the unbound phase is desorbed, and (iii) when both the phases are adsorbed. While in the phases corresponding to the melting type (i) and (iii), the two transitions remain {separated}, for melting type (ii), both the transitions coincide into one transition, represented by an overlapping phase boundary for a considerable range of $g$ and $q$ values, thereby, giving rise to a multicritical line [Fig.~\ref{fig4}]. Intriguingly, the adsorption transition is promoted to first-order in this overlapping region [Fig.~\ref{fig5}(a)].

Adjacent to this overlapping region, and bounded by the lines $g^*=1.3413$ and $q_c=0.2856$ on the other two sides, is a small triangular island (denoted by {\bf a}) {which exhibits qualities} akin to the Borromean phase found in nuclear systems \cite{verdyan2006,comm4}. The {specialty of this} {\bf a} phase {is that it is} not possible when either of the coupling potentials is turned off and exists as a result of the combined effect of the two, even though neither $g$ nor $q$ is strong enough to support an ordered state, individually. This small window of $q$ and $g$ values corresponding to the coinciding phase line facilitates achieving an adsorbed and a bound ({\bf BA}) phase by changing only $g$ or $q$, with the other parameter fixed. Such points (or region) can be crucial for real biological systems since it reduces a multi-parameter system to be controlled by a single parameter. Adsorption in this region follows a similar scaling exponent as of the first-order melting transition. In Fig.~\ref{fig5}(a), the scaling plots are shown for the average number of surface contacts ($n_s$), and the scaling of the fluctuation peaks $C_{s,max}$ in Fig.~\ref{fig5}(a) inset, for $g=1.25$, using $q_c=0.278(4)$ and {${\phi_a=0.96}{\pm 0.02}$} for data collapse. {This is our strongest evidence to support a first-order adsorption}. A similar inter-change of the transition order was previously observed in a theoretical model studying the interplay of helix-coil and adsorption transition in a polymer by Carri and Muthukumar in Ref.~\cite{carri1999}. {The first-order-like adsorption, however, increases the fraction of surface contacts only to $n_s/N\sim 1$ [see Fig.~\ref{fig5}(a)], whereas, the other half increases (with increasing $q$) in a way similar to the continuous transition, but, that is far away from the transition point.}

   {That there is a change in the nature of the adsorption transition} is also evident from the probability distribution of the surface contacts $(P_{n_s})$ {close to} the transition point, e.g., at $g^*=1.25$ and $q=0.2782$ in Fig.~\ref{fig5}(b). {$P_{n_s}$ as found in Fig.~\ref{fig5}(b) is in stark contrast with the critical distribution found for the continuous case} even at chain lengths $N=1000$. Usually, a first-order transition is characterized by a doubly-peaked distribution with a growing depth of the valley in-between, and the gap between the peaks converges to a constant. This valley results from of a $d-1$ dimensional interface separating the {coexisting} phases in the $d$ dimensional system {and incurs an energy penalty while going between the phases. This surface energy, in turn} suppresses the states in between the peaks. It grows exponentially deep in the thermodynamic limit $P\sim \exp(-\sigma L^{d-1})$, where $L$ is the system size and $\sigma$ is related to the surface tension. However, for certain models where the interface separating the phases can be reduced to a point,  the valley  is absent, and the {inter}facial free energy is no longer extensive in $N$, e.g., in our DNA model, the interface between a bound and an unbound segment is a point, in adsorption a point separates the adsorbed and desorbed phases, or the point interface separating the

\begin{figure}
\centering
\hspace{-7cm}(a)\\
\includegraphics[width=\linewidth]{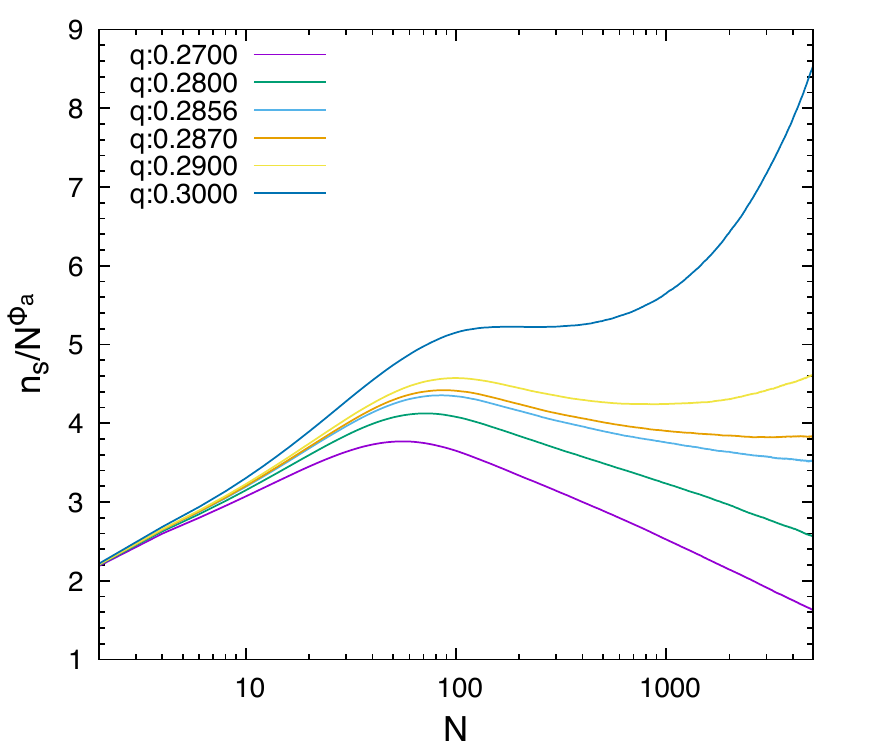}\\
\hspace{-7cm}(b)\\
\includegraphics[width=\linewidth]{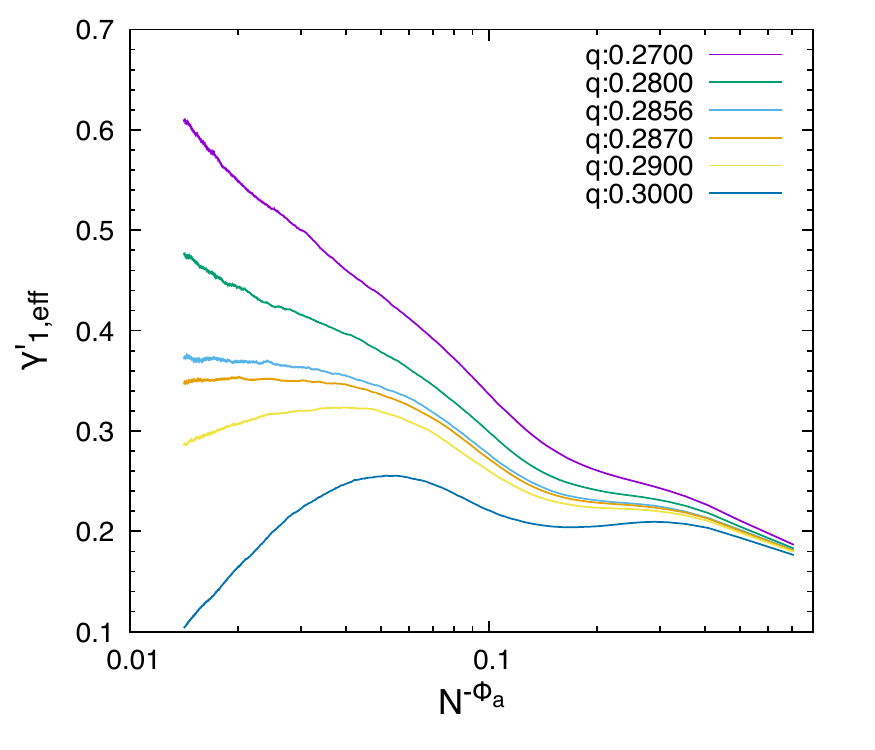}
\caption{(Color online) Model II plots of, (a) average surface contacts scaled by the factor {$N^{\phi_a}$}, and (b)  showing $\gamma_{1,eff}'$ from Eq.~(\ref{SMEQ6}) for different $q$ values. Both plots are for $g=1.17$ and using the scaling exponent $\phi_a=1/2$. }
\label{fig6}
\end{figure}   

\begin{figure}
\centering
\hspace{-7cm}(a)\\
\includegraphics[width=\linewidth]{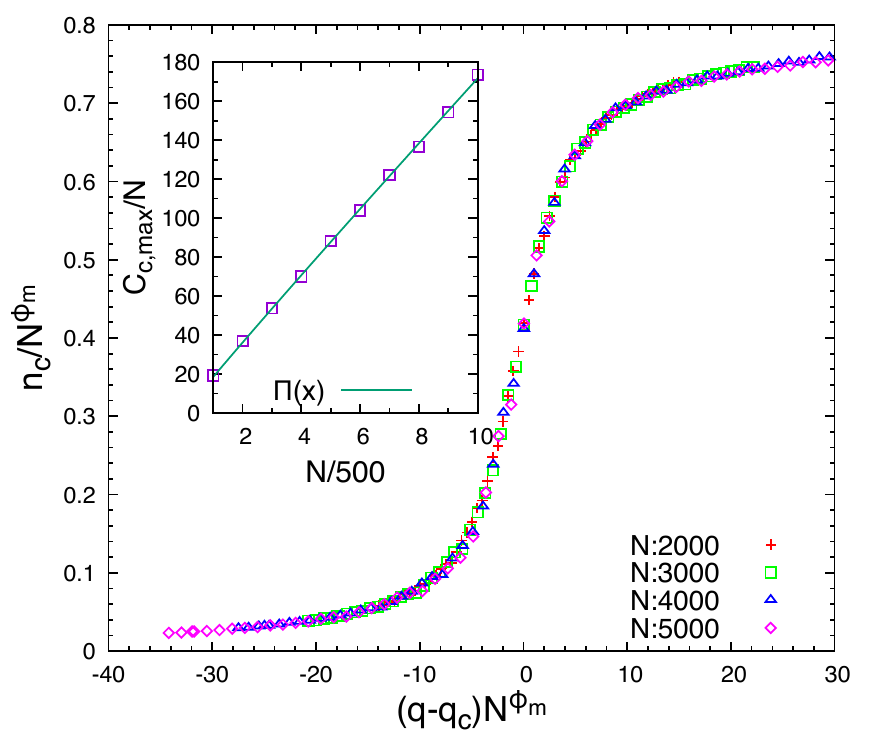}\\
\hspace{-7cm}(b)\\
\includegraphics[width=\linewidth]{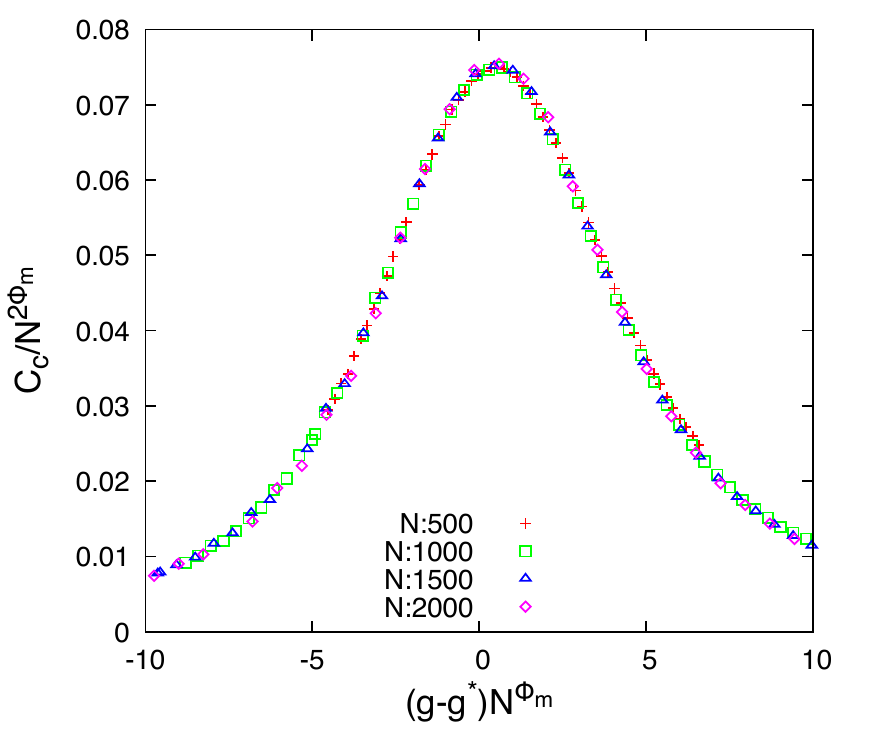}
\caption{(Color online)  (a) Model II scaling plot of the average number of  bp contacts $n_c$, for $g=1.25$, using $\phi_m=0.98{\pm0.01}$ and $q_c=0.278(4)$ for data collapse. ((a) inset) Scaling of the {$n_c$} fluctuation peaks $C_{c,max}$ with $N$. $\Pi(x)\sim x^{0.96\pm0.009}$ is a fit to the data points resulting in $\phi_m=0.98{\pm0.04}$.  (b) Scaling plot of the bps fluctuation ($C_c$) for melting in two-dimensions for $q=7$, using $g^{*}=0.752(6)$ and $\phi_m=0.9{6(3)}$, for chain lengths $N=500$ to $2000$.}
\label{fig7}
\end{figure}

collapsed ferromagnetic phase  from the coiled-paramagnetic phase in the case of a magnetic polymer \cite{garell1999}. {This describes the shape of the distribution in Fig.~\ref{fig5}(b).} {However, we need further simulations with longer lengths and better statistics to rule out any possibility of singular buildups on either arm.}

The melting transition, on the other hand, remains unaffected. To substantiate, we looked at the scaling of the average bp contacts $n_c$ in Fig. \ref{fig7}(a), and the corresponding scaling of the fluctuation peak $C_{c,max}$ with $N$ in Fig. \ref{fig7}(a) inset. The obtained exponent $\phi_m=0.98{\pm0.01}$ from the scaling plots is consistent with the previous value for a first-order melting \cite{causo2000}. Note that, here, the melting is induced by altering the surface interaction for {a} fixed $g$.

Below phase {\bf a}, $q_c$ for an adsorbed phase increases for a small range of $g$ values. The presence of another transition makes it hard to determine $q_c$ from the $C_s$ curves only. Therefore, the $q_c$ values near the confluence point are obtained using Eqs.~\ref{SMEQ5} and \ref{SMEQ6} for chain lengths up to $N=5000$ . While Eq. \ref{SMEQ5} shows a slight increase in $q_c$ [Fig.~\ref{fig6}(a)], Eq. \ref{SMEQ6} do not detect any change [Fig.~\ref{fig6}(b)]. However, since our model has added complexities, e.g., two complementary monomers from different strands can occupy the same lattice site to form a bp which might affect $\mu$ (except for completely bound or unbound state), we believe Eq.~(\ref{SMEQ5}) to give a more reliable estimate of $q_c$.

For {sufficiently high} $q$ values when the system is completely adsorbed ($n_s/2N=1$), the melting transition from {\bf BA} to {\bf UA} phase is two-dimensional.  Since, post-melting, the entropy gain is smaller in the adsorbed phase (two dimensions), compared to the unbound state in the desorbed phase (three dimensions), the bound state in the adsorbed phase is more stable than that in the desorbed phase, leading to a gradual lowering in the threshold $g$, which {eventually} converges to $\lim_{q\rightarrow \infty}  g^*= g_{2d}^*$, where $g_{2d}^*=0.752(3)$ is the two-dimensional melting point. The bp contact fluctuation scaling plot for melting when $q=7$ is shown in Fig. \ref{fig7}(b). {The melting transition remains first-order in 2d.}

A similar argument also applies to the adsorption transition for which the critical adsorption strength $q_c$ decreases and converges to $\lim_{g\rightarrow \infty} q_c = 0.1428$. It is exact{ly half the $q_c$ value of the pure problem} and can be obtained considering that, for model II, even though the {actual contour} length is halved in the bound state ($n_c/N=1$), the energy in the adsorbed phase remains the same. Therefore, the effective adsorbed energy per unit length is doubled, {making it easier to get adsorbed}. {However, the transition is continuous, similar to the pure case.} For example, we estimated the critical adsorption point for $g=5$ to be $q_c=0.1431(5)$.

Unlike model I, the bound state in model II is stabilized in the presence of the adsorbing surface. Although our results from model II are in line with Ref.~\cite{verdyan2006}, qualitatively, we obtain all four possible phases, instead of three, as in \cite{verdyan2006}, where the {\bf UA} phase was absent. Biologically, adsorption-induced stability could be essential to guard DNA native form against thermal fluctuation and external forces. Importantly, adsorption can energetically compensate for the bending of the rigid ds segments, thereby providing an alternative to bubble-mediated bending.

\section{Conclusion} \label{sc6}
To conclude, in this paper, we elucidate the role of  adsorption in modifying the melting transition and vice-versa. Two separate models were considered, which differ in the strength of the interaction with the surface along the ds segments. Such a consideration arises from the speculation that the orientation of the DNA, in conjunction with the nature of the adsorbing surface, could play an important role in determining which of the studied model effectively applies. 

The two models show significant differences: model I shows that the ds structure is mostly destabilized in the presence of an attractive surface, with a small region near the crossing point of the phase lines showing a stable DNA. In contrast to model II, there is no extended region where the phase lines overlap. The findings from this model resemble the result from the experiment performed with DNA hybrids in Ref.~\cite{schreiner2011}. 

On the other hand, model II shows that the ds structure of the DNA  is only stabilized in the presence of an attractive surface with no regions of instability. Although this model is similar to the theoretical model of Ref.~\cite{verdyan2006}, there are significant improvements, such as we consider excluded volume interaction. Moreover, we found the presence of all four possible phases, which is not the case in Ref~\cite{verdyan2006}. Here, we found the presence of an extended region of coinciding phase lines, not present in model I, where adsorption is first-order, and the scaling exponent is similar to that of the melting transition. However, whether this denotes a non-universality in the adsorption transition is yet to be understood. 

{In general, the surface interaction strength for the ds segments could be $\epsilon_{s\text{-}ds}=\alpha \epsilon_{s\text{-}ss}$, where the factor $\alpha\geq 1$. Here, we have studied the two extreme cases $\alpha=1$ and $2$. It will be interesting to see how the two phase diagrams interpolate between model I and II as we continuously vary $\alpha$ from $1$ to $2$, especially if an $\alpha$ exists for which the melting curve remains unaffected by the attractive surface.}

Findings from both models carry biological significance. Our work contributes toward completing the picture by connecting the experimental and theoretical findings, with new results not present in  the previous studies.

\section{Acknowledgments}
D.M. is thankful to Somendra M. Bhattacharjee for valuable discussions. D.M. was supported by the Israel Science Foundation through grant number 1301/17, and the BCSC Fellowship from the Jacob Blaustein Center for Scientific Cooperation. Part of the simulations were carried out on the \emph{Samkhya} computing facility at the Institute of Physics, Bhubaneswar.

\section*{Data Availability Statement}
The data that support the findings of this study are available from the corresponding author upon reasonable request.


\begin{thebibliography}{99}
\bibitem{comm4} In nuclear physics, a Borromean phase is a three-body bound  state in which no two body is individually bound. For DNA adsorption, the two strands and the surface form the three-component system, where any two components are individually capable of forming a bound state.
\end{thebibliography}
\end{document}